\documentclass[aip,jap,twocolumn,showpacs,superscriptaddress,10pt]{revtex4-1}
\usepackage{amsmath,amssymb,graphicx}
\newcommand{\BKFCA}{Ba$_{0.55}$K$_{0.45}$Fe$_{1.95}$Co$_{0.05}$As$_2$}
\newcommand{\SKFCA}{Sr$_{0.6}$K$_{0.4}$Fe$_{1.95}$Co$_{0.05}$As$_2$}

\begin{document}
\title{Enhancement of the upper critical field in codoped
iron-arsenic high-temperature superconductors}

\author{F. Weickert}
\affiliation{Max-Planck-Institut f\"ur Chemische Physik fester
Stoffe, 01187 Dresden, Germany}
\affiliation{Los Alamos National
Laboratory, Los Alamos, New Mexico 87545, USA}
\author{M. Nicklas}
\affiliation{Max-Planck-Institut f\"ur Chemische Physik fester
Stoffe, 01187 Dresden, Germany}
\author{W. Schnelle}
\affiliation{Max-Planck-Institut f\"ur Chemische Physik fester
Stoffe, 01187 Dresden, Germany}
\author{J. Wosnitza}
\affiliation{Hochfeld-Magnetlabor Dresden, Helmholtz-Zentrum
Dresden-Rossendorf, \\ 01328 Dresden, Germany}
\author{A. Leithe-Jasper}
\affiliation{Max-Planck-Institut f\"ur Chemische Physik fester
Stoffe, 01187 Dresden, Germany}
\author{H. Rosner}
\affiliation{Max-Planck-Institut f\"ur Chemische Physik fester
Stoffe, 01187 Dresden, Germany}

\date{\today}

\begin{abstract}
We present the first study of codoped iron-arsenide superconductors
of the 122 family (Sr/Ba)$_{1-x}$K$_x$Fe$_{2-y}$Co$_y$As$_2$ with
the purpose to increase the upper critical field $H_{c2}$ compared
to single doped Sr/BaFe$_2$As$_2$ materials. $H_{c2}$ was
investigated by measuring the magnetoresistance in high pulsed
magnetic fields up to 64\,T. We find, that $H_{c2}$ extrapolated to
$T = 0$ is indeed enhanced significantly to $\approx$\,90\,T for
polycrystalline samples of \BKFCA\ compared to $\approx$\,75\,T for
Ba$_{0.55}$K$_{0.45}$Fe$_2$As$_2$ and BaFe$_{1.8}$Co$_{0.2}$As$_2$ single crystals. 
Codoping thus is a promising way for the systematic optimization of iron-arsenic based superconductors for magnetic-field and high-current
applications.
\end{abstract}


\maketitle

\section{Introduction}

The recent discovery of Fe based superconductors \cite{Kamihara08}
has attracted great interest in the solid-state-physics community.
Up to date, several classes of superconducting Fe based materials
have been found, such as the $RE$OFeAs (1111, $RE$ = rare-earth element),
(Li,Na)FeAs (111),\cite{LiNaFeAs} $A$Fe$_2$As$_2$ (122, $A$ = Ca,
Sr, Ba), \cite{Rotter08a} and FeSe (11) \cite{FeSe} systems.
Superconductivity (SC) at critical temperatures $T_c$ up to 55\,K appears
when the materials are doped appropriately or are pressurized.
Beside the rather high $T_c$, these superconductors have a very high
upper critical field $H_{c2}$. In addition, the small anisotropy of
the 11 and 122 compounds in comparison with 1111-type and especially
cuprate-based high-$T_c$ superconductors, rises reasonable hope,
that they will be available in some years for high magnetic-field
and high-current applications. Another important factor for a
prospective technical implementation are mechanical and processing
properties. In particular, the 122 systems, including the
non-superconducting parent compounds, are metals or semimetals with good mechanical properties for manufacturing.

Concerning the microscopic mechanism, in the compounds of the 1111,
122, and 111 families, the Fe-As layers are considered to be the
structural elements carrying the SC. A charge doping of these layers is most easily accomplished -- as inspired by the
cuprate superconductors -- by an appropriate charge transfer from
the interspacing layer. The tuning of the charge within the
Ba layer by K in Ba$_{1-x}$K$_x$Fe$_2$As$_2$
(\textit{indirect} hole doping) does not change the volume of the
unit cell significantly, but the ratio $c/a$ of the lattice
parameters increases. It was found, that the substitution of Ba by K in the crystal structure of BaFe$_2$As$_2$ leads to a rapid
suppression of the spin density wave (SDW) ordering and to the
emergence of SC.\cite{Rotter08a} The optimum value of
$T_c =$\,38\,K is observed for a K substitution of
40\,{\%}.\cite{Rotter08a} Indirect electron doping failed
until recently, when it was discovered that the substitution of Ca
in CaFe$_2$As$_2$ by La leads to a stabilization of the structurally
collapsed phase and to superconducting signals at temperatures up to
45\,K.\cite{CaLaFe2As2} Concerning upper critical fields, $H_{c2}$ was found to be only
weakly dependent on the direction of the field for K doped BaFe$_2$As$_2$.\cite{Yuan09}
Ba$_{0.55}$K$_{0.45}$Fe$_2$As$_2$, which is closest to optimal doping with a high $T_{c}$ of 32.2\,K, has a maximal
value of 68\,T at $T = 14$\,K ($H \perp c$)\cite{Altarawneh08} and the
extrapolation to zero Kelvin gives an $H_{c2}^0$ of 75\,T.

The \textit{direct} doping of the Fe-As layers by substituting
transition elements on the Fe site \cite{LeitheJasper08b,Sefat08b}
or by replacing arsenic by phosphorous \cite{ShuaiJiang09} was
discovered to be also an option to induce SC in 1111
and 122 compounds. Doping of Co or Ni on the Fe site, which is
equivalent to electron doping, leads to the suppression of the SDW
order and the development of SC as well. In 122 compounds, small cobalt
concentrations of $y = 0.125-0.2$ are sufficient to reach the
highest transition temperature of 18--23\,K for the
SC.\cite{LeitheJasper08b,Sefat08b,Ni08} Also
isoelectronic substitution of Fe by, e.g., Ru was successful to
generate SC,\cite{Schnelle09a} however, the required
concentration is much higher than for electron doping. Hole
doping, e.g., with Mn, has not been successful in neither suppressing the
SDW ordering nor inducing SC.\cite{Kasinathan09a}

The $T_c$ of directly doped 122-type iron-arsenic superconductors is
about a factor of two lower than for those obtained by indirect
doping. This may be understood from the larger disorder in the Fe-As
planes in the former compounds.\cite{LeitheJasper08b} For
BaFe$_{2-y}$Co$_y$As$_2$ the upper critical field for samples close
to optimal doping ($y = 0.2$) was estimated experimentally to be
45\,T at 10\,K,\cite{Yamamoto09} but the steep initial slope of $H_{c2}(T)$ in the temperature-magnetic-field phase diagram suggests an $H_{c2}^{0}$ in the same order of magnitude like for K doped BaFe$_{2}$As$_{2}$.
Subsequent studies on BaFe$_{0.84}$Co$_{0.16}$As$_{2}$ single crystals with same
$T_{c}$ of 22\,K confirm these findings and reveal a critical field $H_{c2}^{0}$ of 55\,T in the limit $T\rightarrow 0$\,K.\cite{Kano09}
Thus, in spite of the lower $T_c$, in-plane
doping seems to be beneficial for obtaining high values of
$dH_{c2}(T)/dT|_{T=T_c}$.

Similar effects are observed for SrFe$_2$As$_2$
doped with K or Co, respectively. Here, the optimal K content is between 0.2 and 0.4
with a corresponding maximum $T_c$ of 38\,K,\cite{Chen08a,Chen08b} however,
for an optimally doped sample a detailed magnetic-field study was
not carried out so far.
In the case of Co doping, the concentration
with the highest $T_c$ of 19.2\,K was estimated to be
0.2 in a study on polycrystalline samples.\cite{LeitheJasper08b} Results of the temperature dependence of the upper critical field
on epitaxial films of the same Co concentration with $T_{c}$\,=17.1\,K reveal a maximum $H_{c2}$ of approximately 47\,T.\cite{Baily09}

In magnetic fields, SC is suppressed mainly by two effects, (i) Pauli limiting, where the Cooper pairs gain more energy than the superconducting condensation energy $\Delta E_{SC}$ via the Zeemann effect and, (ii) by orbital limiting, where the kinetic energy of the electron pairs exceeds $\Delta E_{SC}$. Latter effect can be suppressed by introducing disorder to the system, which systematically enhances the scattering rate. This leads to an increase of $H_{c2}$ and is in particular effective in superconductors with large mean-free-path such as MgB$_{2}$.\cite{Tarantini11} In contrast, Fe based superconductors have a rather small mean-free-path due to a small Fermi velocity and a significant increase of the upper critical field by disorder is unlikely. However, electron and hole doping modifies the shape of the multiple Fermi sheets and can lead to a rise of the upper critical field as well.\cite{Tarantini11}

In the present study, we investigate BaFe$_2$As$_2$ and
SrFe$_2$As$_2$ materials, which were doped simultaneously on the Ba and on the Fe site. Conceptually, our method of codoping is based on the superior
transition temperatures obtained in superconductors of the 122
materials by indirect doping, combined with
an increase of the slope $dH_{c2}(T)/dT|_{T=T_c}$ at $T_{c}$ due to a manipulation of the electronic band structure of the Fe-As layers by direct doping.
The results of our feasibility study strongly
suggest, that by codoping it is possible to push the upper critical
field to higher values. We want to emphasize that no optimization of the concentration levels
has been done so far, but our investigations on \BKFCA\ and
\SKFCA\ establish a straightforward route for an improvement of the superconducting
upper critical field in iron-arsenide superconductors for high-field and high-current applications.

Until recently,\cite{CaLaFe2As2} no electron doping of the Fe-As
layers in 122 systems via charge transfer could be achieved and up
to now no hole doping within the Fe-As layers was successful in
generating SC. We therefore decided to try a
\textit{counter} doping by K (holes) in the interspacing layers and
by Co (electrons) within the Fe-As layers. In order to achieve an
optimum $T_c$, this implies that some overdoping has to be made with
one kind of dopant in order to compensate for the
counter doping. Since $T_c$ might be expected to be low with
large levels of Co doping, we decided to overdope only slightly above the optimal concentration with K
(0.4\,$e^{+}$ + 0.05\,$e^+$ for Ba 122, $\approx 0.35 e^{+}$ + 0.05\,$e^+$ for Sr 122), and to compensate with the same amount of electrons
(-0.05\,$e^-$) with Co.

\section{Experimental Techniques}
\label{sec:exp}

The samples were synthesized by standard powder metallurgical techniques
as described elsewhere\cite{LeitheJasper08b,Kasinathan09a} in an argon-gas
glove-box system. First, stoichiometric blends of reacted starting materials with nominal
compositions Fe$_2$As, Co$_2$As, SrAs, BaAs, and KAs were pressed in
stainless steel dies into cylindrical
pellets without any additional lubricants. Then, these pellets were welded into tantalum containers and sealed under vacuum in quartz ampoules. Heat
treatment at 850\,$^\circ$C for 10 days with up to 4 intermediate regrinding and compaction steps completed the sample preparation. Afterwards, the samples
were cut by a wire-saw and characterized by metallography, also
performed under argon atmosphere, and wave-length dispersive
(WDXS) microprobe analysis. Successful substitution of
alkaline-earth with potassium and iron by cobalt could be
corroborated by quantitative chemical analysis of all investigated
samples. We found, that the exact composition values
Ba$_{0.53(6)}$K$_{0.47(2)}$Fe$_{1.94(6)}$Co$_{0.05(1)}$As$_{1.99(6)}$
and Sr$_{0.6(1)}$K$_{0.4(1)}$Fe$_{1.95(5)}$Co$_{0.06(1)}$As$_{1.96(6)}$ deviate only negligible from the nominal concentrations \BKFCA\ and \SKFCA. The tetragonal lattice
parameters were established to $a$ = 3.9063(2)\,{\AA}, $c$ = 13.365(1)\,{\AA} and $a$ =
3.8918(2)\,{\AA}, $c$ = 12.9095(8)\,{\AA} for the Ba and Sr compound, respectively. Small amounts of
Fe$_2$As as an impurity phase, which grains are randomly dispersed
within the matrix, could also be detected. The material shows a fair
stability towards air and humidity, but starts to degrade after
several weeks.

Resistivity and specific heat were measured between 1.8\,K and
300\,K in a commercial system (PPMS, Quantum Design). The heat
capacity was determined with a thermal-relaxation technique and
resistance with a conventional 4-probe method. Magnetic-susceptibility data were taken around the superconducting phase
transition in a SQUID magnetometer (MPMS, Quantum Design).
The measurement of the magnetoresistance in fields up to 64\,T took place at the Dresden
High Magnetic Field Laboratory (HLD) in a liquid-nitrogen cooled
pulsed-magnet equipped with a $^4$He flow cryostat. The pulsed
magnets at HLD are energized via a 24\,kV capacitor bank. We use a
standard 4-probe AC technique with currents of about 2\,mA and a
measuring frequency of 15\,kHz. The current was applied
parallel to the magnetic-field direction in order to avoid parasitic
Hall-contributions in the results. We observe a small difference in
the magnetoresistance between up and down sweep of the magnet pulse
indicating eddy-current heating, albeit the cross section of the
sample was chosen to be as small as possible ($< 0.2 \times
0.3$\,mm$^2$, see insets in Figs. \ref{fig:Ba4} and
\ref{fig:Sr5}). An estimate of the induced heat in the samples
revealed an increase in temperature of about $\Delta T = 0.5$\,K at
25\,K in the normal conducting state. Nevertheless, we only use data taken
during the up sweep of the magnet pulse to ensure that the indicated
temperatures are correct.

\section{Results and Discussion}
\label{sec:res_dis}

\begin{figure}
\includegraphics[width=90mm]{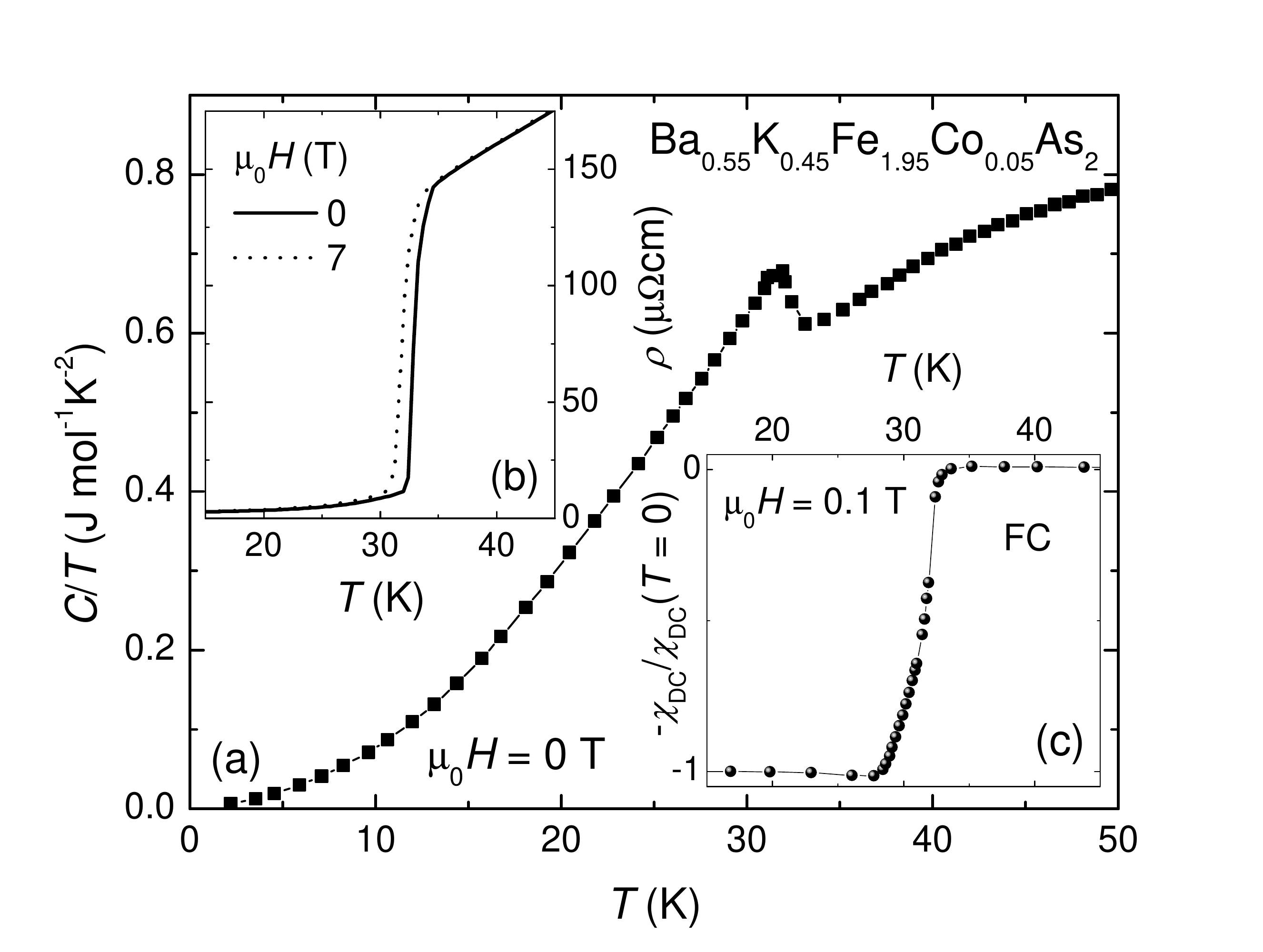}
\caption{Temperature dependence of (a) the specific heat in zero
magnetic field, plotted as $C/T$ vs.\ $T$, (b) the resistivity
$\rho(T)$ in $\mu_{0}H = 0$ and 7\,T, and (c) the magnetic
susceptibility $\chi_\mathrm{DC}(T)$, measured at 0.1\,T in
field-cooled mode and normalized by the low-temperature value of
the \BKFCA\ polycrystalline sample.}\label{fig:Ba1}
\end{figure}

We performed a comprehensive characterization of the samples in order to ensure the good quality of our quinary compounds. 
Metallography on polished surfaces of the sample combined
with WDXS analysis confirm the composition and
good homogeneity. 

The susceptibility data, taken in a field
of 0.1\,T, of \BKFCA\ are shown in Fig.\,\ref{fig:Ba1}(c). A sharp
drop at $T_c$ to large negative values is observed during cooling
the sample in field (FC). The observed onset of the transition is at
$T_{c,\mathrm{on}}$ = 32.5\,K.

Figure\,\ref{fig:Ba1}(a) shows the total specific heat of the same
sample. Specific heat is an excellent tool to judge on the quality
of a superconducting compound since it is a bulk probe. The size
of the step-like specific-heat anomaly at $T_c$ directly measures
the condensation entropy of the superconductor. Moreover, the width
of the step in the present case is predominately determined by the chemical inhomogeneities within the material.
We observe a large and sharp anomaly at $T_c$. The data between 24\,K and 42\,K were fitted with a
lattice term plus a superconducting contribution for a quantitative
analysis. We model the lattice background by a smooth polynomial equation, because no suitable non-superconducting reference sample is available. 
The electronic term can be modeled either with the Bardeen-Cooper-Schrieffer (BCS) theory in the weak-coupling limit, where
the specific-heat jump at $T_{c}$ obeys $\Delta c_{el} = 1.43\gamma_nT_c$, or by the
classical two-fluid model. Latter is a rather good description
for the thermodynamics of a SC with moderately strong
electron-phonon coupling and yields $2\gamma_nT_c$ for $\Delta c_{el}$. For \BKFCA , the
two-fluid model fits better (40\,\% lower least-squares deviation) our specific-heat data,
than the BCS function. The resulting $T_c$ is 32.5\,K, obtained from the midpoint of
the transition, and the value for the reduced specific-heat jump $\Delta
c_{el}/T_c$ can be estimated to 116 mJ\,mol$^{-1}$\,K$^{-2}$. Both, the assumption of moderately
strong coupling as well as the estimated value for $\Delta c_{el}/T_c$ are in agreement with recent
findings in Ref.\ [\onlinecite{Popovich10a}] for
Ba$_{0.68}$K$_{0.32}$Fe$_2$As$_2$, which has a $T_c$ of 38.5\,K. Therefore, we conclude, that the major sacrifice for codoping with 2.5\,\% cobalt is
the lowering of $T_c$ by 6\,K compared to optimal K single doping while the condensation entropy is only
slightly affected. Another quality criterion from specific-heat data is the size of
the residual normal electronic coefficient $\gamma_\mathrm{res}$ in
zero field. Most probably such a contribution indicates
normal conducting parts within the sample, however, the presence of
non-gapped parts of the Fermi surface in certain compositions of
Fe-As superconductors cannot be excluded.\cite{LeitheJasper08b} We
find a particular low $\gamma_\mathrm{res}$ of 2.4 mJ\,mol$^{-1}$\,K$^{-2}$ in our polycrystalline sample. For
comparison, single-crystalline samples of
Ba$_{0.6}$K$_{0.4}$Fe$_2$As$_2$ showed $\gamma_\mathrm{res}$ = 1.2
mJ\,mol$^{-1}$K$^{-2}$, while for
BaFe$_{2-x}$Co$_x$As$_2$ crystals $\gamma_\mathrm{res}$ = 3.7
mJ\,mol$^{-1}$K$^{-2}$ was reported.\cite{Popovich10a,Gofryk10c} The initial Debye temperature of \BKFCA\ is 232.(3)\,K.

The electrical resistivity decreases continuously between room
temperature and 32.5\,K (data not shown), which is characteristic
for metallic behavior. The value slightly above $T_c$ [Fig.\,\ref{fig:Ba1}(b)] of about 135\,$\mu\Omega$\,cm indicates low
impurity concentrations in the sample. Figure\ \ref{fig:Ba1}(b) already
gives an idea about the enormous tolerance of the superconducting state
in \BKFCA\ against magnetic fields. In a field of 7\,T, the critical temperature
is suppressed by only 1.5\,K.

Our second sample, \SKFCA\ (data not shown), displays in the
resistivity and magnetic susceptibility a behavior similar to
\BKFCA, but at lower $T_c$ of 29\,K. The anomaly in the specific heat at the superconducting phase transition is less pronounced in this material, which is no indication for bad sample quality, but typical for Sr 122 iron-arsenides.\cite{LeitheJasper08b} Those compounds have a lower
electronic Sommerfeld coefficient, $\gamma$, indicating a lower electronic density
of states at the Fermi level and, therefore, generally lower condensation entropy.
However, the transition in \SKFCA\ is also considerably broader than for
Ba based material, suggesting some degree of chemical inhomogeneity
in the sample. This observation is supported by the more than 2 times higher resistivity value of 325\,$\mu\Omega$\,cm just above $T_{c}$ in the normal conducting state.

\begin{figure}
\includegraphics[width=100mm]{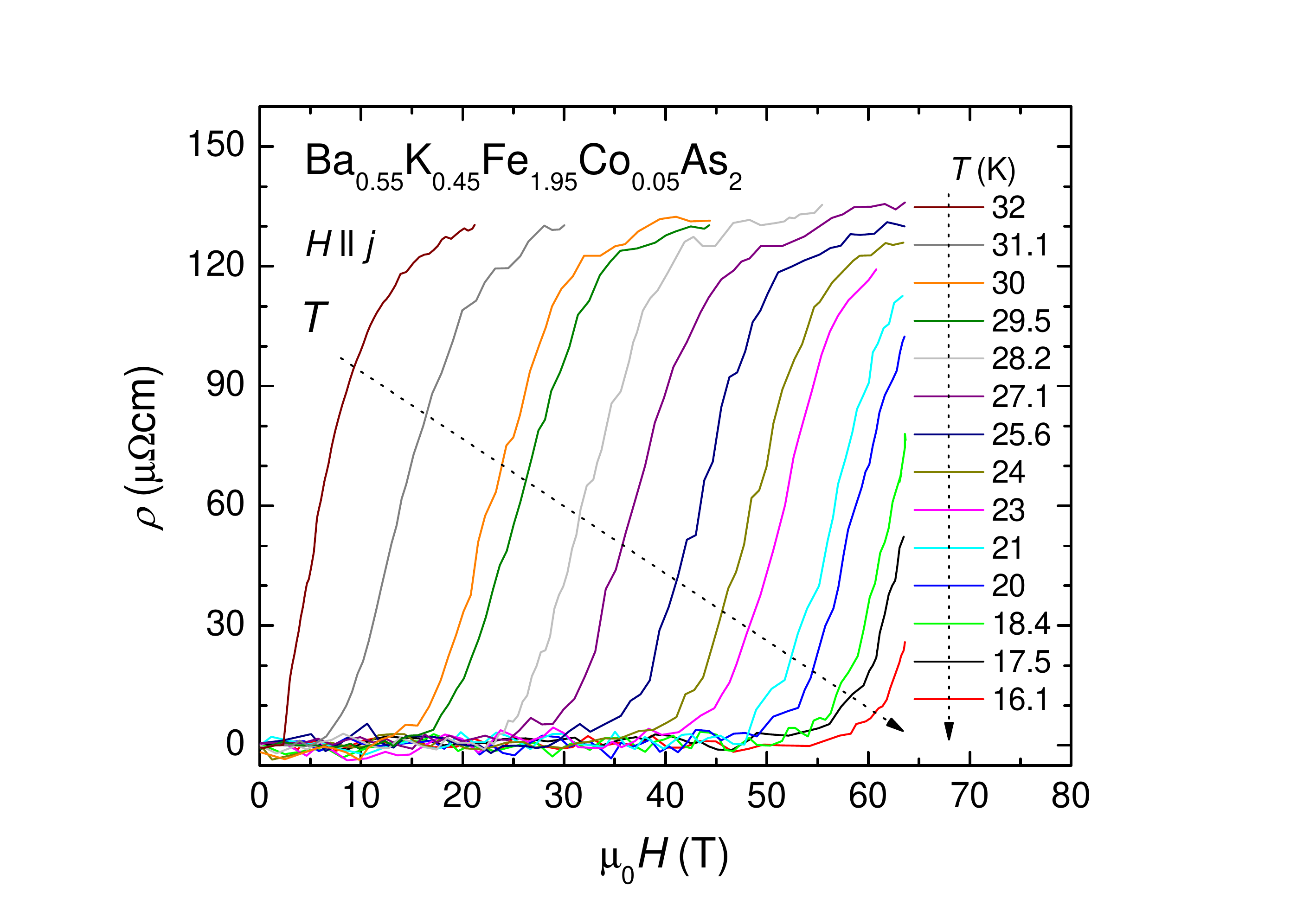}
\caption{(color online) Longitudinal magnetoresistance $\rho(H)$ ($H
\parallel j$) measured up to 64\,T in pulsed magnetic fields for
temperatures between 16.1\,K and 32\,K of a \BKFCA\ polycrystal.}
\label{fig:Ba2}
\end{figure}

\begin{figure}
\includegraphics[width=90mm]{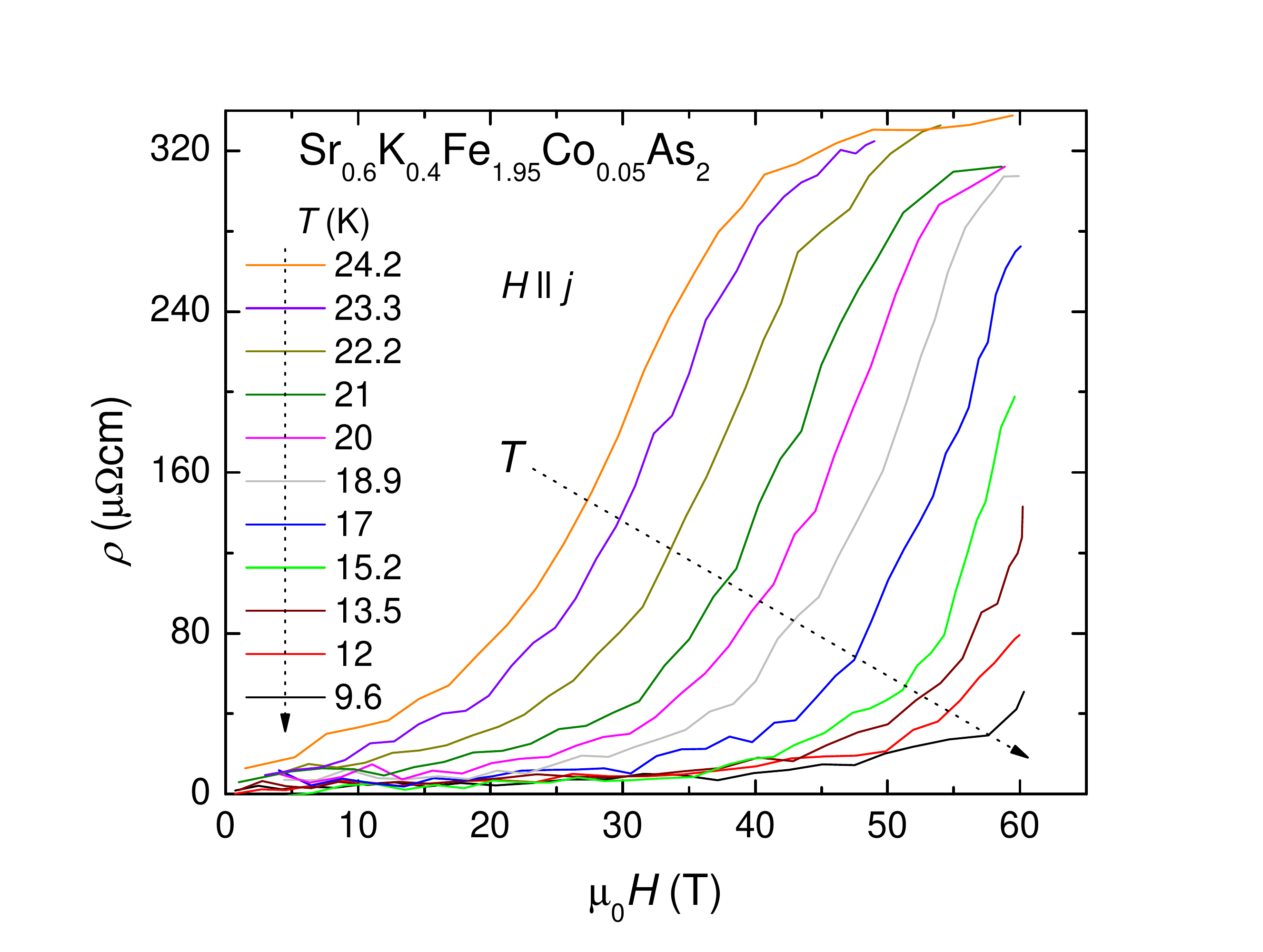}
\caption{(color online) Longitudinal magnetoresistance $\rho(H)$ ($H
\parallel j$) measured up to 60\,T in pulsed magnetic fields for
temperatures between 9.6\,K and 24.2\,K of a \SKFCA\ polycrystal.}
\label{fig:Sr3}
\end{figure}

The high-field magnetoresistance as a function of applied magnetic
field is plotted in Fig.\,\ref{fig:Ba2} for \BKFCA\ and in Fig.\,
\ref{fig:Sr3} for \SKFCA. The superconducting transition in the
first material is rather sharp and the width does almost not change
with decreasing temperature. In contrast, in \SKFCA\ the transition
width is about 30\,T for 23.3\,K just below $T_c$.

The broadening of the superconducting transition for
polycrystalline samples is associated with the anisotropy of the
upper critical field between different crystallographic
directions.\cite{Fuchs09b} In tetragonal systems, such as the
122 compounds presented here, the ratio $\gamma_H = H^{\perp c}_{c2}/H^{\parallel
c}_{c2}$ with $H^{\perp c}_{c2} > H^{\parallel
c}_{c2}$ has to be considered. The magnetoresistance is dominated by the higher critical
field due to a random orientation of crystal grains in the sample.
As long as they are superconducting, the micro crystals oriented
along the direction of the higher critical field ($H \perp c$)
short-circuit the electrical current in the sample. Thus, the
onset of SC in the $\rho(H)$ curves is a good approximation for $H_{c2} \perp c$. The
appearance of finite resistivity, on the other hand, can be used
to identify the lower limit of the distribution of critical fields,
namely $H_{c2} \parallel c$. Nevertheless, one has to keep in mind, that the
anisotropy $\gamma_{H}$ obtained from magnetoresistance measurements
is usually underestimated.\cite{Fuchs09b}

The experimental results in Fig.\,\ref{fig:Ba2} reveal, that the
critical field in \BKFCA\ is quite isotropic with values of
$\gamma_{H}$ around 3.5 near $T_c$, which decrease below 1.5 at
lower temperatures. These ratios are comparable to those observed
for individually K or Co doped single crystals
Ba$_{0.55}$K$_{0.45}$Fe$_2$As$_2$ \cite{Altarawneh08} and
BaFe$_{1.8}$Co$_{0.2}$As$_2$ \cite{Yamamoto09} and emphasize the
essentially three-dimensional character of SC in the
investigated material. For \SKFCA\ the anisotropy is significant
higher and can be estimated to 4.5 at 23.2\,K with decreasing tendency down to
3 at lower temperatures. The values are higher than those observed
for SrFe$_{1.8}$Co$_{0.2}$As$_2$ film samples ($<1.6$),\cite{Baily09} but they
are consistent with recent observations on single crystals of
Sr$_{1-x}$Na$_x$Fe$_2$As$_2$,\cite{Jiao11} where $\gamma_{H}$ varies
between 8 and 2.

\begin{figure}
\includegraphics[width=90mm]{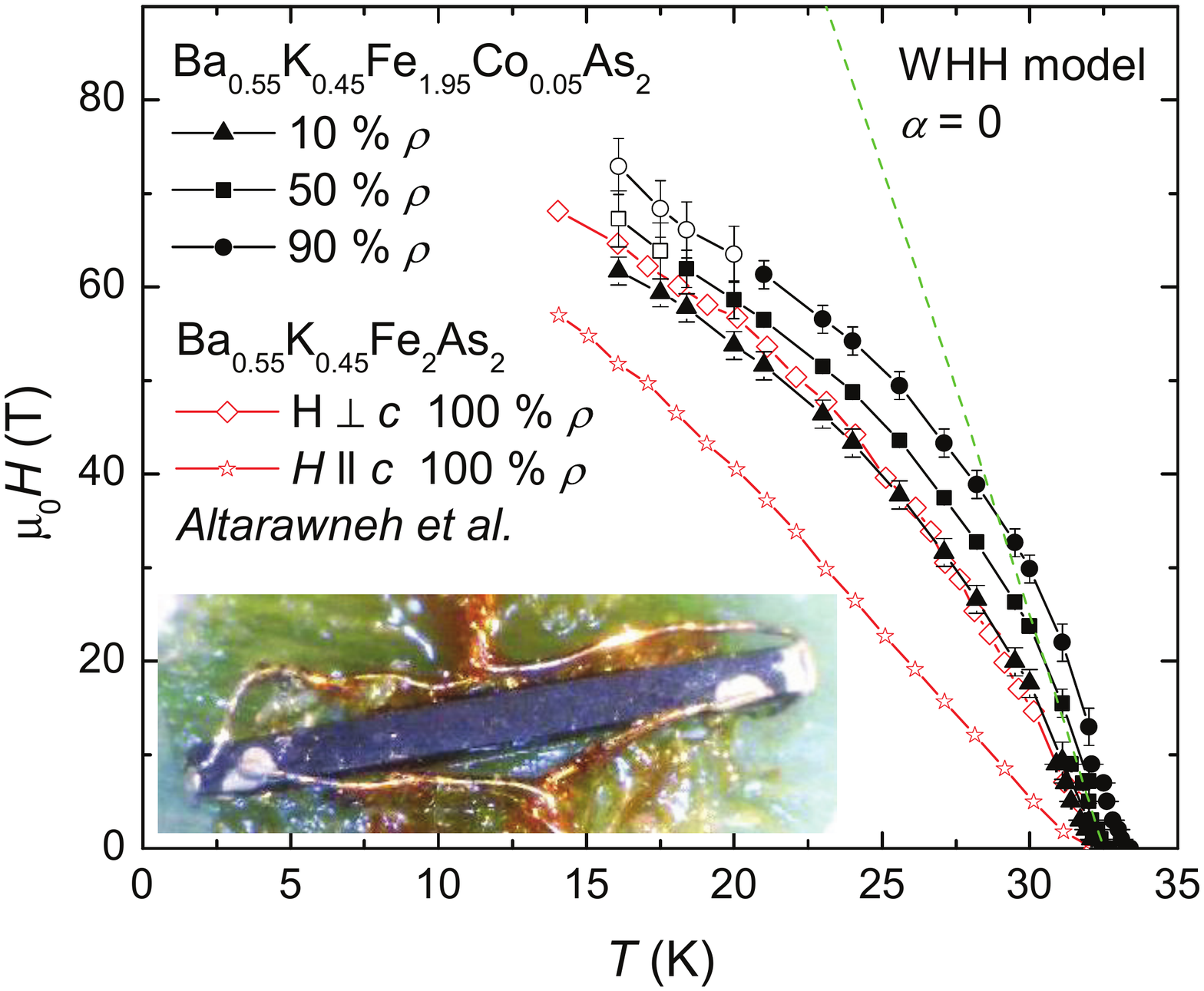}
\caption{(color online) Temperature -- magnetic-field phase diagram
of the superconducting transition in \BKFCA\ estimated by
magnetoresistance measurements. The (black) symbols label $H_{c2}$ for
the 10\,\% (triangles), 50\,\% (squares), and 90\,\% (dots)
criterion of the resistivity. Closed symbols are experimental data,
whereas data marked with open symbols are extrapolated values.
Theoretical expectations of the WHH model without Pauli limiting (Maki parameter $\alpha = 0$) are
indicated by the (green) dashed line. The (red) open stars and diamonds are data
of \textit{Altarawneh et al.}\cite{Altarawneh08} on a single crystal
Ba$_{0.55}$K$_{0.45}$Fe$_2$As$_2$ for magnetic field $H
\parallel c$ and $H \perp c$, respectively. The photo inset shows the
4.5\,mm long polycrystalline sample mounted for the pulsed-field experiment.}
\label{fig:Ba4}
\end{figure}

The magnetic field -- temperature ($H-T$) phase diagram of the
superconducting phase transition is obtained from resistivity data
as a function of temperature in small static magnetic fields and
from the magnetoresistance experiments at constant temperature in
pulsed fields. Before $H_{c2}$ can be estimated from the
magnetoresistance, the data need to be scaled due to a non-negligible
temperature and field dependence of the resistivity in the
normal conducting state, $\rho(T,H)$ for $H > H_{c2}(T)$. Both
effects have been considered to be quadratic in temperature $\Delta
\rho (T) \sim T^2$ and field $\Delta \rho (H) \sim H^{2}$,
respectively.

\begin{figure}
\includegraphics[width=95mm]{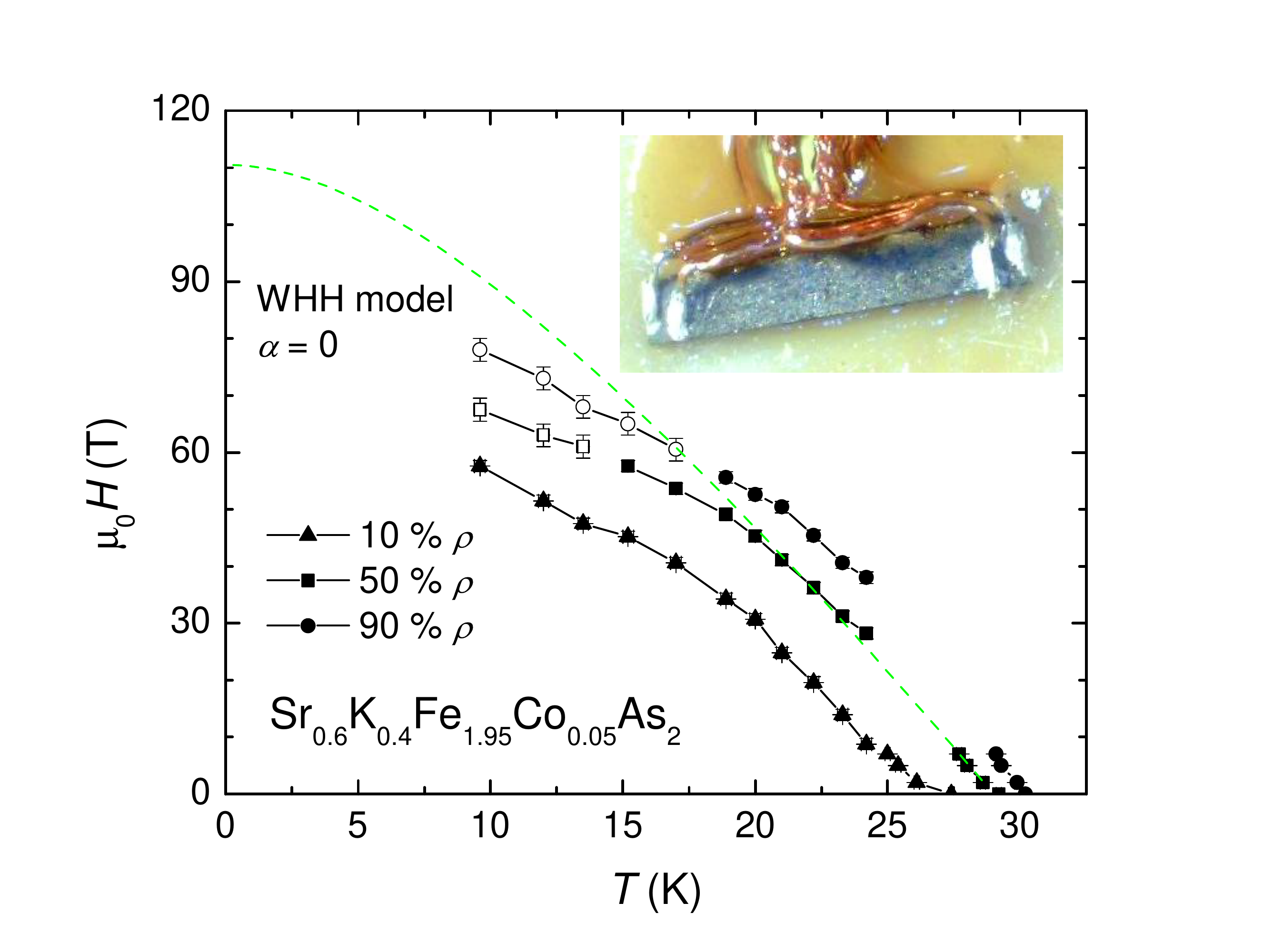}
\caption{(color online) Temperature -- magnetic-field phase diagram
of the superconducting transition in \SKFCA. The symbols
label $H_{c2}$ for the 10\,\% (triangles), 50\,\% (squares), and
90\,\% (dots) criterion of the resistivity $\rho$. Closed symbols
are experimental data, whereas data marked with open symbols are
extrapolated values. Theoretical predictions of the WHH model
without Pauli limiting (Maki parameter $\alpha = 0$) are indicated by the (green) dashed line. The
photo inset shows the 3.5\,mm long polycrystalline sample mounted for magnetoresistance
measurements in pulsed fields.} \label{fig:Sr5}
\end{figure}

We estimate the onset of SC (90\%$ \rho$), the
midpoint of the superconducting transition (50\%$ \rho$), and the
onset of dissipation in the sample (10\%$ \rho$). The data are
displayed in Fig.\,\ref{fig:Ba4} for \BKFCA\ and in Fig.\,
\ref{fig:Sr5} for \SKFCA. We observe in \BKFCA\ a smooth crossover
from data obtained in static fields to those from pulsed-field
experiments. In the case of \SKFCA, pulse-field data are missing
between 25 and 28\,K, because we focused on the high-field region
in the phase diagram during our granted magnet time at the
pulse-field facility. Open symbols in the phase diagrams represent
extrapolated values $H_{c2}^{90\%}$ and $H_{c2}^{50\%}$ for
temperatures, where up to 64\,T (\BKFCA) and up to 60\,T (\SKFCA)
only 10\% of $\rho$ was achieved, assuming a continuous smooth
increase of the magnetoresistance at $H_{c2}$ of similar
shape as for curves at higher temperatures.

One important parameter characterizing the superconducting state,
which is directly linked to the orbital critical field, is the
initial slope $dH_{c2}/dT|_{T=T_c}$. A fitting of our 50\%$\rho$
data reveals a steep slope of $-$10.1\,T\,K$^{-1}$ at $T_c$ for
\BKFCA, which is rather high among all 122 iron-arsenic based
superconductors, but was also found in Ba$_{0.68}$K$_{0.32}$Fe$_2$As$_2$.\cite{Gasparov10} The initial slope for \SKFCA\ instead is with $-$5.5\,T\,K$^{-1}$ a typical value for this class of materials.\cite{Fuchs09b,Putti10}

The Werthammer-Helfand-Hohenberg (WHH) model \cite{Helfand66,
Werthammer66} is commonly used to evaluate the $T=0$ limit of the
upper critical field, $H_{c2}^0$. Accounting only for orbital
effects where the Maki parameter $\alpha$ is 0, the break-down of SC is predicted for
\begin{equation}\label{equ_Hc2}
H_{c2,\textrm{orb}}^0 = 0.69 T_c
\left |\frac{dH_{c2}}{dT} \right |_{T_c}
\end{equation}
and it depends only on the initial slope and the transition
temperature $T_{c}$. We show in both phase diagrams (Figs.\,\ref{fig:Ba4} and
\ref{fig:Sr5}) the expected WWH curves as dashed lines.
The calculated maximal values $H_{c2,\textrm{orb}}^0$ for \BKFCA\ and
\SKFCA\ are 227\,T and 110\,T, respectively. The actual
upper critical fields are much smaller than the
predictions of the WHH model indicating the importance of Pauli
spin-paramagnetism as pair-breaking mechanism for both compounds.
This is in particular valid for the codoped Ba sample.

We compare in Fig.\,\ref{fig:Ba4} our results on \BKFCA\ with data
taken on Ba$_{0.55}$K$_{0.45}$Fe$_2$As$_2$ single crystals
\cite{Altarawneh08} obtained by radio-frequency (RF) penetration-depth
experiments. The comparison shows very clearly that, despite of similar superconducting transition temperatures of about 32.5\,K, the critical field of the codoped sample is significant larger than $H_{c2}$ of Ba$_{0.55}$K$_{0.45}$Fe$_2$As$_2$ in the entire temperature range. We conclude, that extra cobalt doping in a small amount does not reduce the transition temperature, but increases the upper critical field significantly. Note, that we are
comparing our upper-critical-field data using the 90\%$\rho$
criterion with $H_{c2}$ values from the RF penetration-depth
technique, which are basically equivalent to $H_{c2}$ data obtained
from the 100\%$\rho$ rule. This implies, that the difference
between the $H_{c2}(T)$ curves of both materials will increase further when using the same experimental technique.

Finally, the upper critical field in the zero-temperature limit is extrapolated.
Following the curvature of $H_{c2}(T)$ we expect 90\,T for \BKFCA , which
is an increase by more than 15\,\% compared to single doped Ba$_{0.55}$K$_{0.45}$Fe$_2$As$_2$.\cite{Altarawneh08} For \SKFCA\ no high field $H_{c2}(T)$ data on Sr$_{0.6}$K$_{0.4}$Fe$_{2}$As$_{2}$, the reference compound exist. However, a comparison with results on SrFe$_{1.8}$Co$_{0.2}$As$_{2}$\cite{Baily09} suggests a similar enhancement considering the different $T_{c}$'s.

\section{Summary and Outlook}
\label{sec:sum}

To summarize, we have measured the magnetoresistance of
polycrystalline samples of \BKFCA\ and \SKFCA\ between the
superconducting transition temperature and about 10\,K in pulsed
high magnetic fields up to 64\,T in order to estimate the
upper critical field of the superconducting phase transition. We
found that, (i) the codoped sample \BKFCA\ shows in the entire phase diagram significant
larger critical fields than the optimally doped
Ba$_{0.55}$K$_{0.45}$Fe$_2$As$_2$ compound and, (ii) that the
critical field in the 0\,K limit is enhanced by more than 15\%,
despite of the same transition temperature of about 32.5\,K. The results
for the Sr containing 122 material are similar to those of the Ba
compound, whereas $H_{c2}^{0}$ is 80\% higher than for only cobalt doped samples. However, the critical temperatures and the homogeneity of
the Sr samples is lower compared to the Ba system. This is reflected in a significant broader superconducting phase transition.

Our study establishes \textit{codoping} as the route to follow for
enhancing the upper critical field on the iron-arsenide
superconductors with only a minor effect on the transition
temperature. In order to obtain even higher upper critical fields at
$T=0$, a detailed optimization of the K content
with respect to the Co concentration for the highest possible $T_c$
and critical fields is required. We are convinced that our findings
are an important step forward in the development of iron-arsenide
materials for the usage in high magnetic-field and high-current applications.


FW and MN acknowledge financial funding by the MPG Research Initiative:
\textit{Materials Science and Condensed Matter Research at the
Hochfeldmagnetlabor Dresden}. Part of this work was supported
by EuroMagNET II under the EC contract 228043. MN, JW, ALJ, and HR
thank the Deutsche Forschungsgemeinschaft (DFG) for financial
support through SPP 1458.

\end{document}